\documentclass[12pt]{article}
\usepackage{amssymb}
\usepackage{amsmath}
\usepackage{amsthm}
\usepackage{graphicx,psfrag,epsf}
\usepackage{enumerate}
\usepackage{natbib}
\usepackage{tabularx}
\usepackage{url} 
\usepackage{tikz}
\usetikzlibrary{calc,arrows,positioning,decorations.pathreplacing}
\usepackage{bm}
\usepackage{float}
\usepackage{multirow}
\usepackage{threeparttable}
\usepackage{booktabs}
\usepackage{dsfont}
\usepackage{tikz-network}
\newtheorem{prop}{Proposition}
\newtheorem{assum}{Assumption}
\newtheorem{defi}{Definition}

\newtheorem{theo}{Theorem}

\usepackage{color}

\makeatletter
\renewcommand*\env@matrix[1][\arraystretch]{%
  \edef\arraystretch{#1}%
  \hskip -\arraycolsep
  \let\@ifnextchar\new@ifnextchar
  \array{*\c@MaxMatrixCols c}}
\makeatother

\begin{document}

\def\spacingset#1{\renewcommand{\baselinestretch}%
{#1}\small\normalsize} \spacingset{1}


{
  \title{\bf On the Estimation of Peer Effects for Sampled Networks}
      \author{Mamadou Yauck \vspace{.3cm}\\
   Department of Mathematics,\\ Université du Québec à Montréal, \\
    Montréal, QC, Canada H3C 3P8 
}

  \maketitle
}


\bigskip
\begin{abstract}
This paper deals with the estimation of exogeneous peer effects for partially observed networks under the new inferential paradigm of design identification, which characterizes the missing data challenge arising with sampled networks, and based on the idea that two full data versions which are topologically compatible with the observed data may give rise to two different probability distributions. We show that peer effects cannot be identified by design when network links between sampled and unsampled units are not observed. Under the assumption that sampled units report on the size of their network of contacts, and under realistic modeling and topological conditions, the asymptotic bias arising from estimating peer effects with incomplete network data is characterized, and a bias-corrected estimator is proposed. The finite sample performance of our methodology is investigated via simulations.
\end{abstract}

\noindent%
\vfill

\spacingset{1.45} 
\section{Introduction}
\label{sec:intro}
The estimation of peer effects, or the effects of peers' covariate values on one's outcome, has gained an ever growing interest following the groundbreaking works of \cite{Man93} and \cite{bramoulle2009identification}. Recently, novel methodologies have focused on the estimation of peer effects for linear-in-means models under partially observed network data. The most recent papers assume (partial) knowledge about the network formation model, then leverage such information to get consistent estimators for the network parameters (\citealt{chan11, chandrasekhar2016network, boucher2020estimating}). Under the general framework of network reconstruction, every observed subgraph of the population graph is assumed to be a realization from a random network formation process, where the probability of a link formation depends on measured variables and a vector of parameters \citep{chan11}. The main idea is to assume network formation models that guarantee a consistent estimation of the parameter vector given a known sampling process \citep{chandrasekhar2016network}, and to reconstruct the missing links given the observed data.

This paper deals with the estimation of peer effects for partially observed network data under a new paradigm: \textit{design identification}. This new concept formally characterizes the missing data problem arising with sampled networks: two full data versions which are topologically \textit{compatible} with the observed data may give rise to two different probability distributions. The design identification of a given parameter vector, which is intrinsically linked to the sampling procedure, has major implications on the validity of standard inferential procedures. In Section \ref{sec:methodo}, we consider an \textit{exogeneous} peer effects model and show that the parameters cannot be identified under a sampling design in which connections between sampled and unsampled units are not observed. We further investigated the asymptotic bias arising from estimating the peer effects using the incomplete network data under realistic modeling conditions, and under the assumption that sampled units report on the size of their social network contacts within the target population.

	The rest of the paper is organized as follows. In Section \ref{sec:graphst}, we present the graphical structure of the networked population and that of recruitment graphs induced by the sampling process. In Section \ref{sec:methodo}, we present an exogeneous peer effects model in which a unit's outcome may be affected by the neighborhood's mean covariate. Given the outlined design identification problem, we introduce, in Section \ref{sec:ancorr}, addtional assumptions under which we characterize the asymptotic biases of the maximum likelihood estimators (MLE) of the peer effects and propose bias-corrected estimators. We conducted a simulation study in Section \ref{sec:simulations} to assess the finite sample properties of the MLEs and bias-corrected estimators of the peer effects given the observed, incomplete data. 

\section{Population and sample structures}\label{sec:graphst}
This section describes the graphical structure of the network population, details the sampling procedure and formally defines the corresponding recruitment graphs.
 \subsection{The graphical structure of the population}\label{sec:disco}
\begin{figure}[H]
\begin{center}
\includegraphics[scale=.90]{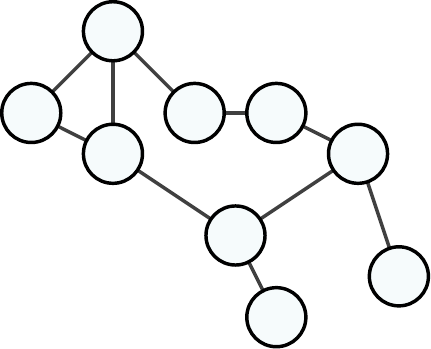}
\end{center}
\caption{Representation of a population graph of 9 vertices.}
\label{fig:dispopnet}
\end{figure}

We consider a population of individuals connected by social ties. This underlying network structure can be represented by a \textit{graph} $G=(V,E)$, where $V$ represents the set of vertices or nodes, with $|V|=N$, and $E$ denotes the set of edges or links shared by nodes. The ordered pair $(u, v) \in E$ denotes an edge, where $u,\,v \in V$. For \textit{undirected} graphs, $(u, v) \in E$ implies $(v, u) \in E$ whereas for \textit{directed} graphs, $(u, v) \in E$ does not imply $(v, u) \in E$. Two vertices are said to be {\it adjacent} if they are linked by an edge. A {\it path} is a sequence of distinct vertices that are adjacent. Two adjacent vertices are called \textit{neighbors}. A graph is said to be \textit{connected} if there is a path from any node to any other node \citep{west2017introduction}.
\begin{assum}{\textbf{(Population graph).}}\label{assum:popgraph}
The population graph is an undirected, connected graph $G=\left(V, E\right)$, where $V$ and $E$ are finite sets, with $|V|=N$.
\end{assum}
Under this assumption, one can reach any vertex from any other vertex in the graph. This is illustrated in Figure \ref{fig:dispopnet}.

%
%
%
%
%
%
Let $x_{j}$ and $y_j$ be trait values for the $j$th vertex, which are assumed independent of the structure of $G$. In Section \ref{sec:methodo}, we assume a model for realizing the random variable $Y_j=y_j$ given $\{x_{j}\}$ and $G$. Let $d_j$ be the number of vertices tied with the $j$th vertex of $G$; this represents the \textit{degree} of the $j$th vertex.

\subsection{Sampling from networks and recruitment graphs}\label{sec:samprocess}
We describe the random node sampling (RNS) as a common strategy for sampling $n$ individuals from the finite population described in Section \ref{sec:disco}. The sampling process, which leverages the underlying network structure to gain more insight from the population, is assumed to navigate across population units' network ties. The RNS procedure is described as follows.
\begin{itemize}
\item[Step 1.] First, randomly select $n$ individuals from the population. This represents the first-level recruitment.
\item[Step 2.] Second, include all social ties between the $n$ selected individuals. This is the second-level recruitment.
\end{itemize}
At the end of the recruitment process, the observed data is $\{x_{j}, y_{j}; i=1,\, j=1, \dots, n\}$; we also observe all connections between recruited individuals. We assume, in addition, that recruited individuals report on their population degrees $\{d_j\}$.

\begin{figure}[H]
\begin{center}
\includegraphics[scale=.90]{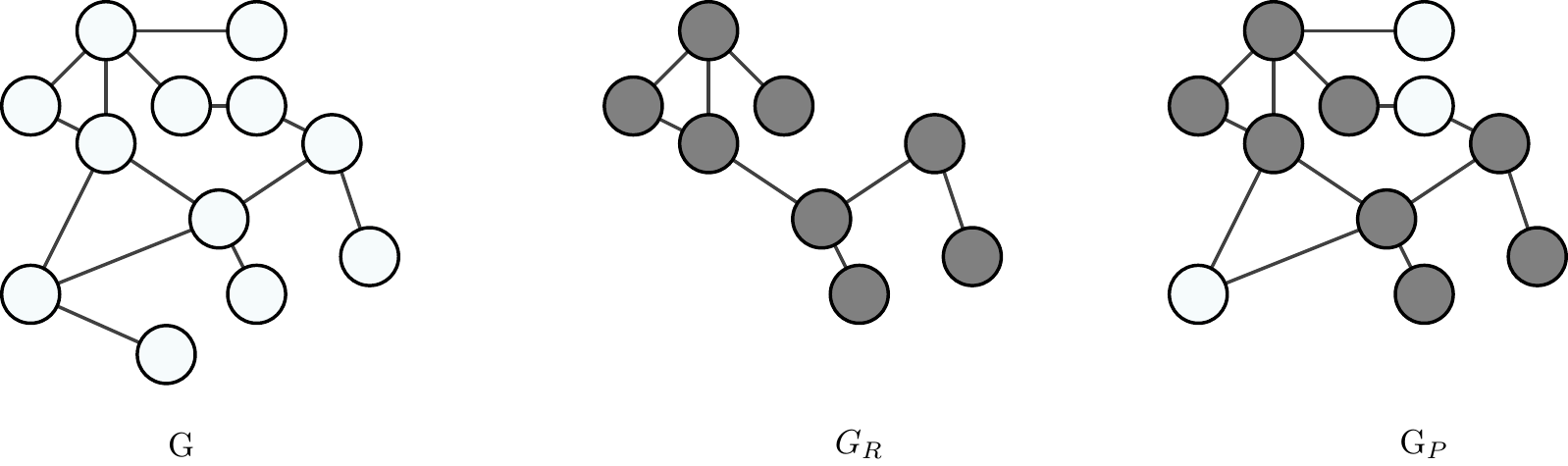}
\end{center}
\caption{Illustration of the population graph $\mbox{G}$, the recruitment graph $G_{R}$ and the  population-induced subgraph $\mbox{G}_P$.}
\label{fig:RDSgraphs}
\end{figure}

Now consider the sampling technique described above and the underlying population network from which sampling took place. A graph $G_{*}=(V_{*},E_{*})$ is a \textit{subgraph} of $G=(V,E)$ if and only if $V_{*}\subseteq V$ and $E_{*}\subseteq E$. First, we define the RNS recruitment subgraph with respect to the population graph $G$. 
\begin{defi}{\textbf{(Recruitment subgraph).}}\label{def:recsubgraph}
 The recruitment subgraph is an undirected graph $G_{R}=(V_{R},E_{R})$, where $V_{R}$ is the set of recruited individuals, and $E_R=\{ (i,j): i \in V_T,\,j \in V_T,\,\, \text{and}\,\, (i,j) \in E\}$ is the set of edges in the population graph $\mbox{G}$ connecting vertices in $V_R$.
\end{defi}
It follows that the RNS recruitment data is $\{x_{j}, y_{j}, d_j, G_R;\, j \in V_R\}$. Now let $V_U=\{i \notin V_R: \exists \,j \in V_R \,\,\text{and}\,\, (i,j) \in E \}$ denote the set of unrecruited individuals linked to at least one individual in the recruitment graph $\mbox{G}_R$ and let $E_U=\{ (i,j): i \in V_R,\,j \in V_U\,\, \text{and}\,\, (i,j) \in E\}$ be the set of edges linking  individuals in $V_U$ to individuals in $V_R$. We define the subgraph induced by $V_U \cup V_R$.
 \begin{defi}{\textbf{(Population-induced subgraph).}}\label{def:recpopsubgraph}
The population-induced subgraph is the undirected graph $G_{P}=(V_P,E_P)$ with $V_P=V_T\cup V_U$ and $E_P=E_R\cup E_U$.
\end{defi}
An example of the population graph $\mbox{G}$ from which sampling took place and its corresponding recruitment subgraphs are illustrated in Figure \ref{fig:RDSgraphs}. 

\section{Model specification and identification}\label{sec:methodo}
In this section, we consider a popular network model in which an individual's outcome may be affected by their covariate and their neighbors' mean value for the covariate. This model is defined over the fully observed population network. We then introduce and define the notion of design identification, which formalizes the missing data problem when the underlying population network is only partially observed. 
\subsection{Model specification}
The target population's individual-level data is given by $\{G, \bm{\mathcal{D}}\}$, where $\bm{\mathcal{D}}=\{y_{j}, x_{j}; j=1, \dots N\}$, and $y_{j}$ ($x_{j}$) is the outcome (the covariate) of the $j$th individual. The elements of the $N\times N$ matrix of social relationships are $(s_{jk})_{j,k}$, where $s_{jk}=1$ if the $j$th individual and the $k$th individual share a tie, that is if $(j, k) \in E$, $s_{jk}=0$ otherwise and $s_{jj}=0$. Consider a model in which an individual's outcome may be affected by the value of their covariate and by the neighbors' mean value for the covariate:
\begin{equation}\label{eq:netmod}
y_{j}=\beta_0+ \beta_1x_j+ \beta_2 d_j^{-1}\sum_{k=1}^{N} s_{jk}x_{k}+\epsilon_{j},
\end{equation}
where $\epsilon_{j} \sim N(0, \sigma^2_{\epsilon})$. The parameter $\beta_2$ captures exogeneous peer effects, i.e., the influence of peers' characteristics on the outcome of an individual. Model (\ref{eq:netmod}) corresponds to a less parameterized version of the classical peer effect model \citep{Man93}. 

Inference for the parameter vector $(\beta_0, \beta_1, \beta_2, \sigma^2_{\epsilon})$ is straightforward when the underlying population network is fully observed. The next section investigates the issues surrounding valid inference for the parameter vector when the network is partially observed.

\subsection{Design identification}\label{subsec:designidenti}
Under the sampling design considered in Section \ref{sec:samprocess}, the observed data is $\{\mbox{G}_R, \bm{\mathcal{D}}_R\}$, where $\bm{\mathcal{D}}_R=\{y_{j}, x_{j}; j=1, \dots n\}$. This section investigates challenges regarding statistical inference for the parameter vector of model (\ref{eq:netmod}) when the full data $\{\mbox{G},\,\bm{\mathcal{D}}\}$ is partially observed. Let $\bm{y}=\left(y_{1},\dots,y_{n}\right)$ and $\bm{x}=\left(x_{1},\dots,x_{n}\right)$. Also, let $\tilde{\bm{x}}=\left(x_{1},\dots, x_{n}, x_{n+1},\dots, x_{n+u}\right)$, where $\{n+1,\dots,n+u\}$ represents the set of unrecruited individuals who share connections with recruited individuals. Finally, let $\bm{\theta}=(\beta_0, \beta_1, \beta_2, \sigma^2_{\epsilon})$ be the parameter vector for model (\ref{eq:netmod}) and $p(\bm{y}|\bm{x}, G;\bm{\theta})$ denote the conditional distribution of $\bm{y}$ given $\bm{x}$ and a graph $G$, for the parameter vector $\bm{\theta}$.

\begin{prop}\label{prop:extrapmod1}
One has: $p(\bm{y}|\bm{x}, G_S;\bm{\theta})=p(\bm{y}|\tilde{\bm{x}}, G_P;\bm{\theta})$.
\end{prop}
The proof follows from the observation that, for each $j \in V_R$,
$$
\sum\limits_{\substack{k \in V: (k, j) \in E}}x_{k}=\sum\limits_{\substack{k \in V_P: (k, j) \in E}}x_{k}.
$$
Proposition \ref{prop:extrapmod1} implies that, for model (\ref{eq:netmod}), inference about $\bm{\theta}$ given $G_S$ and $\bm{x}$ is the same as inference about $\bm{\theta}$ given $G_P$ and $\tilde{\bm{x}}$. Thus, the validity of inference can be assessed using $\{G_P,\,\bm{\mathcal{D}}_P\}$ instead of $\{G,\,\bm{\mathcal{D}}\}$. Next, we define the topological conditions under which $\{\mbox{G}_P, \bm{\mathcal{D}}_P\}$ could be defined as a full data version of the observed data $\{\mbox{G}_R, \bm{\mathcal{D}}_R\}$. This is the notion of \textit{compatibility}.

\begin{defi}{\textbf{(Compatibility).}}\label{def:Tcompat}
$\{\mbox{G},\,\bm{\mathcal{D}}\}$ is compatible with the observed data $\{\mbox{G}_R, \bm{\mathcal{D}}_R\}$, if $(i)$ $V_R \subseteq V$, $(ii)$ $E_R \subset E$ and $(iii)$ $\bm{\mathcal{D}}_{R} \subset \bm{\mathcal{D}}$. Let $\mathcal{C}_{(G_R, \bm{\mathcal{D}}_{R})}$ denote the set of full data $\{\mbox{G},\,\bm{\mathcal{D}}\}$ that are compatible with the observed data $\{\mbox{G}_R,\,\bm{\mathcal{D}}_{R}\}$. 
\end{defi}
The main implication of the definition is that any two full data distributions will give rise to the same observed data distribution. We now give a formal definition of identification.

\begin{defi}{\textbf{(Identification).}}\label{def:d-ident}
The parameter $\bm{\theta}$ is said to be identified given $\{\mbox{G}_R,\,\bm{\mathcal{D}}_{R}\}$ if
$$
\{G^{'}_P,\,\bm{\mathcal{D}}_P^{'}\},\,\{G^{''}_P,\,\bm{\mathcal{D}}^{''}_P\} \in \mathcal{C}_{(\mbox{G}_R, \bm{\mathcal{D}}_{R})}\,\,\, \text{implies} \,\,\, p(\bm{y}|\tilde{\bm{x}}, \mbox{G}^{'}_P;\bm{\theta})= p(\bm{y}|\tilde{\bm{x}}, \mbox{G}^{''}_P;\bm{\theta}).
$$
\end{defi}
Identification is not achieved when there exist two full compatible data whose distributions are different. Thus, identification in this context is different from structural identification, which depends on the parameterization of the model. The lack of identification in this context implies that parameters associated with network or group effects cannot be consistently estimated without additional assumptions about the structure of the population graph.

The concept of design identification is crucial in formalizing the missing data problem for the following reason. Some authors have approached regression for partially observed network data through the lens of a measurement error problem, without refering to the inderlying topology of the unobserved population graph. This is important because not all specifications of the covariate for $\beta_2$, given a set of configurations for the population graph, would result in the covariate being mismeasured. As an example, let the covariate for $\beta_2$ be $T_p(\mbox{G})-T_a(\mbox{G})$, where $T_p(\mbox{G})=\sum_{j<k<\ell} \mbox{1}(s_{jk}+s_{j\ell}+s_{\ell k}\geq 2)$ is the number of connected triples and $T_a(\mbox{G})=\sum_{j=1}^N\sum_{k=1}^N\sum_{\ell=1}^N s_{jk}s_{j\ell}s_{\ell k}$ is the number of triangles in the population graph, and assume a sampling design in which the observed graph is $\mbox{G}_R$. Suppose we know the following about the structure of the population graph: recruited individuals $j, k \in E_R$ share ties with two unrecruited individuals each,  $N=\sum_{i=1}^m n_i+4$, unrecruited individuals $\ell, r, s, t \in V_U$ are such that $d_r=d_s=d_t=d_u=2$, and $(r, t), (r, u) \notin E$. In this scenario, all configurations for $G$ will result in an error-free covariate for $\beta_2$.  

The next proposition addresses the design identifiability of $\bm{\theta}$ when model (\ref{eq:netmod}) is fitted to $\{\mbox{G}_R, \bm{\mathcal{D}}_R\}$. 
\begin{prop}\label{prop:identmod1}
Given $\{\mbox{G}_R,\bm{\mathcal{D}}_{R}\}$, the parameter vector $\bm{\theta}$ is not identified.
\end{prop}
The proof is given in the Appendix. This result implies that consistent estimation of the parameter vector is not guaranteed given the observed data. In the next section, we investigate the asymptotic bias arising from estimating $\bm{\theta}$ via maximum likelihood using the observed RNS data $\{G_R, \bm{\mathcal{D}}_R\}$, and propose a bias-corrected estimator under realistic modeling assumptions and under an additional condition about the topology of $G$.

\section{Model-based analytical corrections}\label{sec:ancorr}
The missing data induce a measurement error in the covariate for $\beta$, thus creating bias for the MLE of $\bm{\theta}$ as $\mathbb{E}[\epsilon_{j}|\bm{x}]\neq 0$ for some $j \in V_R$ where $\mathbb{E}[.]$ denotes expectation. In this section, we treat the estimation of $\bm{\theta}$ as a classical measurement error problem and propose novel bias-corrected estimators. First, we make additional assumptions under which we characterize the asymptotic bias arising from estimating $\bm{\theta}$ via maximum likelihood using $\{G_R, \bm{\mathcal{D}}_R\}$. We then propose a bias-corrected estimator, which is consistent under these assumptions. Let $d^R_{j}$ be the number of ties that the $j$th individual share with other individuals within the recruitment graph $G_R$. The results of this section are derived using the conditions defined in the following assumption.
\begin{assum}\label{assum:regcond}
Let (\ref{eq:netmod}) be the underlying data generating model, with $\mathbb{E}[\epsilon_{j}|\tilde{\bm{x}}]=0$ for all $j$. We assume:
\begin{eqnarray}
 \label{eq:cond1}\mbox{plim}~ n^{-1}\sum_{j}(x_{j}-\bar{x})^2&=&\sigma_x^2,\,\, \text{with}\,\, 0<\sigma^2_x<\infty, \\
\label{eq:cond2}\mbox{plim}~ n^{-1}\sum_{j\neq k}(x_{j}-\bar{x})(x_{k}-\bar{x})&=&0, \\
\label{eq:condeps}  \mbox{plim}~ n^{-1}\sum_{j}x_{j}\epsilon_j&=&0,\\
\label{eq:cond3} \lim_{n\to\infty} d^R_n/d_n&=&w_R.
\end{eqnarray}
\end{assum}
Equation (\ref{eq:cond1}) is a regularity condition while (\ref{eq:condeps}) ensures asymptotic independence between the error and the covariate. Condition (\ref{eq:cond3}) is that of uniform convergence for the ratio of the observed degree in the recruitment graph to the true degree in the population graph; it directly links, through network degrees, a structural part of the observed graph topology to that of the unobserved population graph. Given $\{G_R, \bm{\mathcal{D}}_R\}$, condition (\ref{eq:cond2}) guarantees that $\beta_1$ can be consistently estimated; it also ensures sign-consistency for the MLE of $\beta_2$. The next proposition formally characterizes the bias of the MLE of $\beta_2$ when model (\ref{eq:netmod}) is fitted to $\{G_R, \bm{\mathcal{D}}_R\}$.
 \begin{prop}\label{prop:betacorrT}
Consider model (\ref{eq:netmod}) and let $\hat\beta_{2,R}$ be the estimator of $\beta_2$ obtained by maximum likelihood given $\{G_R, \bm{\mathcal{D}}_R\}$. Under conditions (\ref{eq:cond1}), (\ref{eq:cond2}) and (\ref{eq:cond3}) defined in Assumption \ref{assum:regcond}, 
\begin{equation*}
\mbox{plim}~\hat\beta_{2,R}=w_R\beta_2.
\end{equation*}
\end{prop}
The proof is given in the Appendix. Proposition \ref{prop:betacorrT} suggests constructing a new estimator $\hat\beta^c_{2,R}=w_R^{-1}\hat\beta_{2,R}$. It is easy to show that this estimator is consistent, thus asymptotically unbiased, under conditions (\ref{eq:cond1}), (\ref{eq:cond2}) and (\ref{eq:cond3}). Note that the rescaling factor $w_R$ depends on individuals' true network degrees $\{d_n\}$, which are assumed reported in the RNS recruitment procedure. It follows from Proposition \ref{prop:betacorrT} that $\mbox{plim}~\hat\beta_{2,R}\leq \beta_2$ since $w_R\leq 1$. The finite sample performance of the proposed estimator will be investigated via simulations in Section (num.). Empirically, the quantity $\sum_{j=1}^n d^{-1}_j/\sum_{j=1}^n (d^R_j)^{-1}$ can be used as a scaling factor for a sample of size $n$, as justified by the next Proposition.
 \begin{prop}\label{prop:scalingfactor}
Under the uniform convergence condition (\ref{eq:cond3}), one has 
$$
\lim_{n\to\infty} \frac{\sum_{j=1}^n 1/d_j}{\sum_{j=1}^n 1/d^R_j}=w_R.
$$
\end{prop}
The proof is given in the Appendix. The next theorem characterizes the asymptotic normality of the corrected estimator $\hat\beta^c_{2,R}$.
\vspace*{2mm}
 \begin{theo}\label{theo:betacorrTNorm}
Consider model (\ref{eq:netmod}) and let $\hat\beta^c_{2,R}$ be the corrected estimator of $\beta_2$ given $\{\mbox{G}_R, \bm{\mathcal{D}}_R\}$. Under conditions (\ref{eq:cond1}), (\ref{eq:cond2}) and (\ref{eq:cond3}) defined in Assumption \ref{assum:regcond}, assuming that $\mathbb{E}[(Y_j-\bar{Y})^3]<\infty$, and under the additional condition
\begin{equation}\label{eq:asynormcond}
\left(\sum_{j=1}^n|x_j-\bar{x}|^3\right)^2=o\left(\sum_{j=1}^n(x_j-\bar{x})^2\right)^3,
\end{equation}
one has:
\begin{equation}\label{eq:asynorm}
\sqrt{n} \left(\hat\beta_{2,R}^c-\beta_2 \right)\rightarrow N\left(0, \frac{w^{-2}_R\sigma^2_{\epsilon}}{\sum_{j=1}^n(x^*_j-\bar{x^*})^2}\right),
\end{equation}
where $x^*_j=\frac{1}{d^R_j}\sum_{k=1}^{n}s_{jk}x_{k}$ and $\bar{x^*}=\frac{1}{n}\sum_{j=1}^n x^*_j$.
\end{theo}
The proof is given in the Appendix. Theorem \ref{eq:asynorm} implies that we can obtain, under the aforementioned assumptions, consistent confidence intervals for the peer effects $\beta_2$.

\section{Simulation study}\label{sec:simulations}
The goal of this study is to assess the accuracy and precision of the maximum likelihood estimator of $\beta_2$ given the observed data, $\hat\beta_{2,R}$, and that of the corresponding bias-corrected estimator $\hat\beta_{2,R}^C$, as well as the coverage for the 95\% confidence interval of $\beta$. We computed standard errors using the observed Fisher information matrix. Confidence intervals for the bias-corrected estimators were obtained by dividing the confidence intervals limits for the MLEs of $\beta_2$ by the corresponding rescaling factors.

We generated a network of size $N=10^3, 10^4$ for which the probability of a tie between two nodes is $p=1\%,\,3\%$; $p$ represents the \textit{density} of the network, or the number of observed connections over the number of possible connections. We used a sample fraction of $f=20\%, 80\%$ and reported all connections between sampled units in each case. Model (\ref{eq:netmod}) was simulated as follows. A continuous covariate $x$ was generated from a normal distribution with mean $3$ and standard deviation $1.5$; we set the parameter vector to $(\beta_0, \beta_1,\beta_2,\,\sigma^2_{\epsilon})=(0,\,1,\,1.5,\,1)$. The model was then fitted to the incomplete data $\{\mbox{G}_{R}, \bm{\mathcal{D}}_{R}\}$. We ran $10,000$ simulation repetitions for each combination of simulation parameters and computed the bias and the root mean squared error of $\hat\beta_{2,R}$ and $\hat\beta^C_{2,R}$, and the coverage for the 95\% confidence intervals of $\beta_2$.

The results are displayed in Table \ref{table:simulation_CVGE}. The estimator $\hat\beta_{2,R}$ exhibits substantial bias for small sample fractions, which then decreases as the sample fraction increases. Note that the sign of the biases align with findings of Section \ref{sec:ancorr} that $\hat\beta_{2,R}$ is sign-consistent and conservative under the conditions of Assumption \ref{assum:regcond}. The bias-corrected estimator $\hat\beta^C_{2,R}$ shows negligible bias, with increasing precision, across all sample sizes and sample fractions.

\begin{table}[H]
\caption{ Relative bias (RB) and root mean squared error (RMSE) of the MLE $\hat\beta_{2,R}$ and bias-corrected estimator $\hat\beta_{2,R}^C$ of $\beta_2$, and coverage (CI) for the 95\% confidence interval of $\beta_2$, for increasing population size ($\mbox{N}$), network density ($p$) and sample fraction ($\mbox{f}$).}
\begin{center}
\setlength\extrarowheight{-4.5pt}
\begin{tabular}{lc c c  c ccc ccc} \hline
\multicolumn{6}{r}{$\mbox{f}=20\%$}&&&&{$\mbox{f}=80\%$}\\
    \cline{5-7}     \cline{9-11} 
{$\mbox{N}$} &{$p$}& & {Estimator}& {RB}&{RMSE}&{CI}  &&{RB}&{RMSE}&{CI}\\ \hline
\multirow{6}{1em}{$10^3$}  &  \multirow{2}{1em}{$1\%$}& & $\hat\beta_{2,R}$ & -0.76&  1.15&0.00  &  &-0.19&0.30&0.63\\
&& & $\hat\beta_{2,R}^C$ & 0.02 &  0.49&0.96  &  & $<$0.01 &0.12&0.96\\\\
&   \multirow{3}{1em}{$3\%$}&   & $\hat\beta_{2,R}$ & -0.78&1.18&0.00  &  &-0.19&0.32 &0.81\\
&&   & $\hat\beta_{2,R}^C$  & -0.01&0.71&0.94  &  &$<$0.01 &0.16&0.94 \\
\hline
\multirow{6}{1em}{$10^4$}  &  \multirow{2}{1em}{$1\%$}&   & $\hat\beta_{2,R}$ &-0.79 &  1.18&0.00   &  & -0.19&0.30&0.43 \\
&&   & $\hat\beta_{2,R}^C$ &-0.02 &  0.35& 0.95  &  &$<$0.01&0.08& 0.95\\\\
&  \multirow{3}{1em}{$3\%$}&   & $\hat\beta_{2,R}$& -0.79& 1.19&0.00  & & -0.20&0.33&0.74 \\
&&   & $\hat\beta_{2,R}^C$& 0.01&0.57&0.94 & & $<$0.01&0.16&0.95 \\
\hline
\end{tabular}
\end{center}
\label{table:simulation_CVGE}
\end{table}

\section{Discussion}
The validity of inference for statistical network models, when the population graph is assumed fully observed, has been extensively studied in the literature. However, when the population network is not fully observed, standard inferential methods may not be valid. By characterizing the problem as one of identification driven by the study design, this paper demonstrates that, in some classical network models, valid inference for peer effects, or the influence of your peers on an outcome of interest, cannot be achieved when the true graph of social relationships is only partially observed. Under some realistic assumptions, we characterized the asymptotic bias of the MLE of the peer effects parameter by treating the lack of identification as a classical measurement error problem. Using information on individuals' reported population network degrees, we proposed a bias-corrected estimator; the asymptotic normality of the proposed estimator was then established. Finally, we empirically showed that the bias-corrected estimator have good finite sample properties.

It is important to point out that this work focused on linear regression models. We expect the identification problem to persist in a generalized linear model setting, where the outcomes may follow binomial or Poisson distributions, but it is unclear whether the analytical results could hold. This will be the subject of further investigation.

Even though identification is often not achievable, the observed RDS data may reveal useful information about the parameters that are not identifiable. This is commonly referred to as \textit{partial identification} (\citealt{Manski03, Romano10, Moon12}). Under partial identification, it may be possible to derive the identification region, defined by the set of values of the target parameters which are compatible with the observed data. However, the feasibility of this approach in a regression setting remains an open question. This will be the subject of future research.

\appendix

\section*{Appendix}

\subsection*{Proof of Proposition \ref{prop:identmod1}}
Let $\{G^{'}_P,\,\bm{\mathcal{D}}^{'}_P\} \in \mathcal{C}_{(G_R, \bm{\mathcal{D}}_R)}$ and $\{G^{''}_P,\,\bm{\mathcal{D}}^{''}_P\} \in \mathcal{C}_{(G_R, \bm{\mathcal{D}}_R)}$, with $G^{'}_P=(V^{'}_P,E^{'}_P)$ and $G^{''}_P=(V^{''}_P,E^{''}_P)$. Let $j \in V_R$ and $\ell \in V_R$ be recruited individuals. Suppose $j$ and $\ell$ share each (at least) a tie with an unrecruited individual. Suppose $G^{'}_P$ and  $\bm{\mathcal{D}}^{'}_P=(y_j, x^{'}_1, \dots, x^{'}_{n+u}; j=1,\dots,n)$ are such that edges $(j, u_1) \in  E_S^{'}$, $(\ell, u_2) \in E_P^{'}$ and $x_{u_1} \neq x_{u_2}$, where $u_1$ and $u_2$ are two unrecruited individuals. Also, suppose $\{G^{''}_P,\,\bm{\mathcal{D}}^{''}_P\}$ is the same as $\{G^{'}_P,\,\bm{\mathcal{D}}^{'}_P\}$ except $(j, u_2) \in  E_P^{''}$, $(\ell, u_1) \in E_P^{''}$. Let $\bm{S}^{'}$ and $\bm{S}^{''}$ represent adjacency matrices for $G^{''}_P$ and $G^{'}_P$, respectively.
Let $\mu^{'}_r=\beta_0+\beta_1 x_r+\beta_2(d_r)^{-1}\sum \limits_{\substack{k=1}}^Ns^{'}_{rk}x_k$, where $d_r=\sum_{k \sim r}s_{rk}$, $r \in V_R$ and $k \in V_P$. The distribution of $\{G^{'}_P,\,\bm{\mathcal{D}}^{'}_{P}\}$ is
$$
p(\bm{y}|G^{'}_P;\bm{\theta}) =(2\pi \sigma^{2}_{\epsilon})^{-n/2}\mbox{exp}\left\lbrace -\frac{1}{2\sigma^{2}}\sum \limits_{\substack{r=1}} ^n \left(y_r-\mu^{'}_r \right)^2 \right\rbrace.
$$
The distribution of $\{G^{''}_P,\,\bm{\mathcal{D}}^{''}_{P}\}$ can be written similarly. One has $p(\bm{y}|G^{'}_P;\bm{\theta}) =p(\bm{y}|G^{''}_P;\bm{\theta}) $ if and only if
\begin{eqnarray}\label{proofeq1}
\nonumber \sum_{r=1}^n (d_r)^{-1}\sum \limits_{\substack{k \sim r }}s^{'}_{rk}x_k &=& \sum_{r=1}^n (d_r)^{-1}\sum \limits_{\substack{k \sim r }}s^{''}_{rk}x_k \,\,\, \Leftrightarrow \\
\nonumber \sum_{r \in \{j,\,\ell\}} (d_r)^{-1}\sum \limits_{\substack{k \sim r }}s^{'}_{rk}x_k &=& \sum_{r \in \{j,\,\ell\}} (d_r)^{-1}\sum \limits_{\substack{k \sim r }}s^{''}_{rk}x_k \,\,\, \Leftrightarrow \\
 (d_j)^{-1}(x_{u_1}-x_{u_2})&=& (d_{\ell})^{-1}(x_{u_1}-x_{u_2}).
\end{eqnarray}
Since $x_{u_1}-x_{u_2} \neq 0$, Equation (\ref{proofeq1}) reduces to $d_j=d_{\ell}$, which is not guaranteed by the design. This completes the proof.

\subsection*{Proof of Proposition \ref{prop:betacorrT}}
We assume, without loss of generality, that $\beta_0=\beta_1=0$. Let $x_{R,j}^d=(d^R_j)^{-1}\sum_{k\sim j}x_{k}$, $\bar{x}_R^d=n^{-1}\sum_{j}x_{R,j}^d$, $x_{j}^d=d_j^{-1}\sum_{k\sim j}x_{k}$, $\bar{x}^d=n^{-1}\sum_{j}x_j^d$ and $\bar{\epsilon}=n^{-1}\sum_{j=1}^n \epsilon_j$. The maximum likelihood estimator of $\beta_2$ is
\begin{eqnarray}\label{eq:MLEbeta}
\nonumber \hat\beta_2&=&\frac{n^{-1}\sum_{j=1}^n (x_{R,j}^d-\bar{x}_R^d)(y_j-\bar{y})}{n^{-1}\sum_{j=1}^n (x_{R,j}^d-\bar{x}_R^d)^2}\\
 \nonumber  &=& \frac{\beta_2 \times n^{-1}\sum_{j=1}^n (x_{R,j}^d-\bar{x}_R^d)(x_{j}^d-\bar{x}^d)+n^{-1}\sum_{j=1}^n (x_{R,j}^d-\bar{x}_R^d)(\epsilon_j-\bar{\epsilon})}{n^{-1}\sum_{j=1}^n (x_{R,j}^d-\bar{x}_R^d)^2}.
\end{eqnarray}
For each recruited individual $j$, let $\{x^j_1, \dots, x^j_{d^R_j}\}$ be the neighbors' covariates within $G_R$. For any reordering of this set, we can write:
$$
\sum_{j=1}^n (x_{R,j}^d-\bar{x}_R^d)^2=\frac{1}{(d^R_n)^2}\left\lbrace \sum_{\ell=1}^{d_n}\sum_{j=1}^n (x_{\ell}^j-\bar{x}_{\ell})^2+\sum_{u\neq v} \sum_{j=1}^n (x_{u}^j-\bar{x}_{u})(x_{v}^j-\bar{x}_{v}) \right\rbrace,
$$
where $\bar{x}_{\ell}=(d_n^R)^{-1}\sum_{j=1}^{n} x^j_{\ell}$. Further,
$$
\sum_{j=1}^n (x_{R,j}^d-\bar{x}_R^d)(x_{j}^d-\bar{x}^d)=\sum_{j=1}^n (x_{R,j}^d-\bar{x}_R^d)^2+\sum_{j=1}^n (x_{R,j}^d-\bar{x}_R^d)(x_{j}^U-\bar{x}^U),
$$
where $x_{j}^U=(d_j-d_j^U)^{-1}\sum_{k, j \in V_U: k\sim j}x_{k}$ and $\bar{x}^U=(|V_U|)^{-1}\sum_{j=1}^{|V_U|}x_{j}^U$.
It follows, from conditions (\ref{eq:cond1}), (\ref{eq:cond2}) and (\ref{eq:condeps}) of Assumption \ref{assum:regcond}, that 
$$
\mbox{plim}~\hat\beta_{2,R}=\beta_2 \times \lim_{n\to\infty} \frac{d^R_j}{d_j}.
$$
Condition (\ref{eq:cond3}) completes the proof.

\subsection*{Proof of Proposition \ref{prop:scalingfactor}}
It is straightforward to show that $\sum_{j=1}^{\infty} 1/d^R_j=\infty$. Now fix $\epsilon>0$. Since $\lim_{n\to\infty} d^R_n/d_n=w_R$, there exists $n_0(\epsilon)$ such that for $n>n_0$,
$$
w_R-\epsilon<\frac{1/d_n}{1/d^R_n}<w_R+\epsilon.
$$
Multiplying by $1/d^T_n$ on both sides of the inequality, with straightforward developments, leads to:
$$
\frac{(w_R-\epsilon)\sum_{j=n_0+1}^n 1/d^R_j+\sum_{j=1}^{n_0} 1/d_j}{\sum_{j=1}^n 1/d^R_j}<\frac{\sum_{j=1}^n 1/d_j}{\sum_{j=1}^n 1/d^R_j}<\frac{(w_R+\epsilon)\sum_{j=n_0+1}^n 1/d^R_j+\sum_{j=1}^{n_0} 1/d_j}{\sum_{j=1}^n 1/d^R_j}.
$$
We first show that the bound on the left hand side converges to $w_R-\epsilon$. One has
\begin{eqnarray}
\nonumber \lim_{n\to\infty} \frac{(w_R-\epsilon)\sum_{j=n_0+1}^n 1/d^R_j+\sum_{j=1}^{n_0} 1/d_j}{\sum_{j=1}^n 1/d^R_j}&=&\lim_{n\to\infty} \frac{(w_R-\epsilon)\sum_{j=n_0+1}^n 1/d^R_j}{\sum_{j=1}^n 1/d^R_j}  \\
\nonumber &=&w_T-\epsilon.
\end{eqnarray}
Using the same approach for the bound on the right hand side of the inequality, we can easily show that 
$$
 \lim_{n\to\infty} \frac{(w_R+\epsilon)\sum_{j=n_0+1}^n 1/d^R_j+\sum_{j=1}^{n_0} 1/d_j}{\sum_{j=1}^n 1/d^R_j}=w_T+\epsilon.
$$

\subsection*{Proof of Theorem \ref{theo:betacorrTNorm}}
We first note that, if condition (\ref{eq:asynormcond}) for $\{x_j\}$ and $\bar{x}$, it will be the same for $x^*$ and $\bar{x}^*$, where $x^*_j=\frac{1}{d^R_j}\sum_{k=1}^{n}s_{jk}x_{k}$ and $\bar{x^*}=\frac{1}{n}\sum_{j=1}^n x^*_j$. Next, we use a special case of the central limit theorem for independent, non-identically distributed random variables.
\begin{theo}\label{theo:CLTnoniid}
Let $Z_1, Z_2, \dots$ be i.i.d with $\mathbb{E}(Z_i)=0$, $\mbox{Var}(Z_i)=\sigma^2>0$, and $\mathbb{E}(Z_i^3)=\delta<\infty$. Then
$$
\frac{\sum_{j=1}^n d_{nj}Z_j}{\sigma \sqrt{\sum_{j=1}^n d^2_{nj}} }\rightarrow N(0, 1),
$$
provided
$$
\left(\sum_{j=1}^n|d_{nj}|^3\right)^2=o\left(\sum_{j=1}^n d^2_{nj}\right)^3.
$$
\end{theo}
The result follows from Theorem \ref{theo:CLTnoniid}, with $d_{nj}=x^*_j-\bar{x}^*$ and $Z_j=Y_j-\bar{Y}$.

\end{document}